\documentclass[aps,pre,reprint,amsmath,amssymb,floatfix]{revtex4-2}

\usepackage[utf8]{inputenc}
\usepackage[T1]{fontenc}
\usepackage[english]{babel}
\usepackage{graphicx}
\usepackage{xcolor}
\usepackage{microtype}
\usepackage[colorlinks=true,linkcolor=blue,citecolor=blue,urlcolor=blue]{hyperref}

\newcommand{\includefigure}[2][]{%
  \IfFileExists{#2}{\includegraphics[#1]{#2}}{\fbox{\parbox{0.88\linewidth}{\centering Missing figure file: \detokenize{#2}}}}%
}

\newcommand{\panel}[1]{\par\vspace{2pt}{\small (#1)}}

\makeatletter
\g@addto@macro\UrlBreaks{\do\-}
\makeatother
\newcommand{\doilink}[2]{\href{https://doi.org/#1}{#2}}

\begin{document}
\makeatletter
\let\auto@bib\@empty
\let\auto@bib@innerbib\@empty
\makeatother

\title{Gradient-based inverse design of free-energy landscapes with diffusion models}

\author{Eli Zick}
\author{Dan Mendels}
\email[Contact author: ]{danmendels@technion.ac.il}
\affiliation{The Wolfson Department of Chemical Engineering, Technion -- Israel Institute of Technology, Haifa 32000, Israel}

\date{\today}

\begin{abstract}
Free-energy surfaces govern the populations of metastable states and the barriers that control transitions between them, making their direct optimization a central challenge in molecular and materials design. In this work, we introduce Gradient-Based Free Energy Surface Optimization (GB-FESO), a proof of concept for an inverse design framework that uses a trained conditional diffusion model as a differentiable surrogate for the ensemble distribution. After training, the diffusion model is frozen, and the conditioning variables defining the system are optimized so that the generated ensemble reproduces a prescribed target free-energy surface. The optimization is carried out by backpropagating a distribution-level loss, based on the squared maximum mean discrepancy or a kernel density estimate of the Kullback--Leibler divergence, through a deterministic diffusion sampling trajectory. We first validate GB-FESO on one-dimensional Gaussian ensembles, demonstrating that both continuous and relaxed discrete conditioning variables can be optimized to recover target distributions. We then apply the method to a four-particle Lennard--Jones toy peptide and a system of two five-bead chains in contact, both exhibiting multiple metastable conformational states. In these more physically motivated settings, GB-FESO successfully optimizes the interaction parameters to reproduce target free-energy landscapes in the majority of test cases, with optimization performed either in the full internal-coordinate space or in a reduced collective-variable representation. These results establish GB-FESO as a promising first step toward an ensemble-level inverse design framework for molecular systems with prescribed thermodynamic and kinetic behavior.
\end{abstract}

\maketitle

\section{Introduction}
\label{sec:intro}

Recent progress in machine learning has expanded the methodological basis of molecular engineering, particularly through models that learn directly from large-scale sequence and structural data~\cite{Butler2018MLMolecularMaterials,SanchezLengeling2018InverseMolecularDesign,Rives2021ProteinLM,Shteingolts2026StructureOnly,SanchezGonzalez2020GraphNetworks}. Within this landscape, Denoising Diffusion Probabilistic Models (DDPMs)~\cite{diffusionOG} have emerged as a useful generative framework for protein ensembles. Unlike classical machine learning approaches that typically produce point predictions, diffusion models are trained to learn the target ensemble distribution itself, through reverse denoising dynamics that map noise to samples from that distribution. Landmark models such as AlphaFold have demonstrated the power of deep learning for high-accuracy protein structure prediction, while recent ensemble-oriented models such as BioEmu extend this progress toward scalable generation of protein equilibrium conformational ensembles~\cite{AlphaFold,BioEmu}. In parallel, diffusion-based protein design models such as RFdiffusion, Chroma, FoldingDiff, and FrameDiff have shown that generative denoising procedures can sample realistic and designable protein backbones, often under structural or functional constraints~\cite{RFdiffusion,Chroma,FoldingDiff,FrameDiff}.

Since many properties of chemical systems are governed by conformational dynamics and relative stability, the ability of DDPMs to learn ensemble distributions makes them a promising approach for inverse engineering. Representative examples from prior work include ExEnDiff~\cite{ExEnDiff}, which guides a pretrained diffusion-based ensemble sampler with experimental observables as physical priors, enabling generation of conformational ensembles that better satisfy measurements while approximating the underlying Boltzmann distribution. ExEnDiff was further shown to capture mutation-induced conformational changes and to help identify useful collective variables for protein ensembles. In a related direction, Minimum-Excess-Work Guidance~\cite{MinimumExcessWorkGuidance} formulates the guidance of pretrained score-based generative models using a thermodynamic-work-inspired regularization principle, allowing sparse restraints or experimental observables to steer sampling while limiting distortion of the learned ensemble. STARLING~\cite{STARLING}, by contrast, is a latent-space DDPM for intrinsically disordered regions that generates coarse-grained conformational ensembles directly from sequence. In addition to ensemble prediction, STARLING introduces ensemble-aware latent representations that support biophysical similarity search and enable latent-space, ensemble-first sequence design, where candidate sequences are optimized to match a target conformational behavior.

These studies illustrate the growing role of diffusion models as ensemble generators and inverse-design tools. At the same time, most current applications use these models primarily to produce structures, ensembles, or sequences that satisfy a prescribed condition. A complementary approach is to treat the trained generative model itself as a differentiable surrogate of the ensemble distribution and optimize the underlying design variables (e.g., interaction energies, chemical composition, or amino acid sequence) such that a desired ensemble distribution is achieved. This shifts the inverse-design problem from targeting individual configurations, such as a protein's native state, to tuning the parameters that govern the full conformational ensemble.

Such an approach is particularly relevant because the behavior of many physical systems is often governed not by a single static structure but by their underlying free-energy surface (FES), which can be expressed as:
\begin{equation}
F(\mathbf{s}) = -k_B T \ln P(\mathbf{s}) + C ,
\label{eq:free-energy}
\end{equation}
where \(P(\mathbf{s})\) is the equilibrium probability distribution of the system over a set of system variables \(\mathbf{s}\), \(k_B\) is the Boltzmann constant, \(T\) is the temperature, and $C$ is an arbitrary constant. The FES governs the population of a system's metastable states and the free-energy differences between them, thus also dictating the overall transition rates. From an engineering perspective, tailoring the FES therefore provides a powerful route to controlling the function of a broad class of systems for which optimization of a single native-like structure alone would be insufficient. Example applications include catalysis, ligand recognition~\cite{Stenstrom2024LigandRecognition}, antibody engineering~\cite{Spoendlin2025AntibodyFlexibility}, and allosteric regulation~\cite{Nussinov2026Allostery}, which frequently depend on how accessible specific conformations are, and on how readily the system can transition to and from them~\cite{Frauenfelder1991}.

Directly sampling the FES is often a challenging and computationally demanding task because barrier-crossing configurations typically have very low Boltzmann probabilities. As a result, conventional simulations tend to become trapped in metastable basins, requiring exceedingly long simulation times to adequately sample transitions and achieve convergence of ensemble populations~\cite{TorrieValleau1977,LaioParrinello2002,Henin2022}. Enhanced-sampling methods can mitigate this limitation, for example, by biasing selected collective variables, thus modifying the effective energy landscape, but their accuracy often depends on identifying variables that capture the relevant slow modes sufficiently well. Thus, engineering the FES of systems of interest, or predicting how structural or chemical changes (e.g., mutations in biomolecules) reshape it, remains a nontrivial design problem~\cite{Henin2022,Medaparambath2026,Alex2025,AbramsBussi2014,Dan2022,Dan2023}.

Motivated by this perspective, we introduce Gradient-Based FES Optimization (GB-FESO), a design framework in which the objective is not merely structural accuracy or the optimization of specific properties, but the controlled optimization of the ensemble distribution itself, for example, the relative free-energy differences between a system's metastable states, the overall shape of these states, or the barriers separating them. GB-FESO employs a conditional diffusion model as a differentiable generator of conformational ensembles. By leveraging the differentiability of this model, it enables the optimization of system design variables such that the generated ensemble approaches a prescribed target distribution. In this way, the optimization is driven by an ensemble-level objective rather than by agreement between individual generated and target conformations.

\section{Methods}
\label{sec:methods}

To enable FES tailoring, we employ a conditional DDPM that serves as a differentiable mapping from a design condition $\mathbf{c}$, representing a set of system-defining properties (e.g., its physical structure, chemical composition or amino acid sequence), to an ensemble of samples that collectively define the corresponding FES. After training on a collection of state (conformational) ensembles spanning a diverse set of systems within the domain of interest, the parameters of the generative neural network are frozen. At this stage, the conditional diffusion model is no longer used only to sample ensembles for a prescribed condition; instead, it is treated as a differentiable surrogate for the condition-dependent ensemble distribution. Inverse design is then performed by optimizing the condition vector $\mathbf{c}$ itself, with gradients obtained by backpropagating an ensemble-level loss through the deterministic sampling trajectory, so that the generated ensemble approaches the desired FES, or equivalently, the desired ensemble statistics.

\subsection{Diffusion models}
\label{sec:diffusion}

DDPMs learn to generate samples by reversing a gradual noising process. Starting from a clean sample $\mathbf{x}_0$, the forward noising process adds Gaussian noise over $T$ steps according to
\begin{equation}
q(\mathbf{x}_t|\mathbf{x}_{t-1}) =
\mathcal{N}(\mathbf{x}_t; \sqrt{1-\beta_t}\mathbf{x}_{t-1}, \beta_t\mathbf{I})
\label{eq:ddpm1}
\end{equation}
where $\{\beta_t\}_{t=1}^T$ is a predefined variance schedule and $\mathbf{I}$ is the identity matrix. Defining $\alpha_t=1-\beta_t$ and $\bar{\alpha}_t=\prod_{s=1}^t \alpha_s$, the marginal distribution at an arbitrary step admits the closed form
\begin{equation}
q(\mathbf{x}_t|\mathbf{x}_0)=
\mathcal{N}(\mathbf{x}_t;\sqrt{\bar{\alpha}_t}\mathbf{x}_0,(1-\bar{\alpha}_t)\mathbf{I})
\label{eq:ddpm2}
\end{equation}
which can be equivalently written as
\begin{equation}
\mathbf{x}_t =
\sqrt{\bar{\alpha}_t}\mathbf{x}_0+
\sqrt{1-\bar{\alpha}_t}\boldsymbol{\epsilon},
\qquad \boldsymbol{\epsilon} \sim \mathcal{N}(\mathbf{0},\mathbf{I})
\label{eq:ddpm3}
\end{equation}

This formulation is useful because it allows noisy samples at any step to be obtained directly from $\mathbf{x}_0$, without explicitly simulating all previous diffusion steps. As $t$ increases, the signal component is progressively reduced while the noise component becomes dominant, so that $\mathbf{x}_T$ approaches an isotropic Gaussian distribution.

The generative model is then trained on the reverse process, which starts from Gaussian noise and iteratively removes noise until a clean sample is obtained. Specifically, a neural network $\epsilon_\theta(\mathbf{x}_t,t,\mathbf{c})$ is trained to predict the noise $\boldsymbol{\epsilon}$ that was added to $\mathbf{x}_0$ to obtain $\mathbf{x}_t$. Training at every diffusion step is important because the amount of corruption varies along the diffusion trajectory, requiring the model to adapt its denoising behavior to different noise levels. In practice, training is commonly performed using a simple mean-squared-error objective on the predicted noise~\cite{diffusionOG,SD}:
\begin{equation}
\mathrm{MSE} = \frac{1}{n}\sum_{i=1}^{n}(\epsilon_i - \hat{\epsilon}_i)^2
\label{eq:mse}
\end{equation}

\begin{figure*}[t]
    \centering
    \includefigure[width=\textwidth]{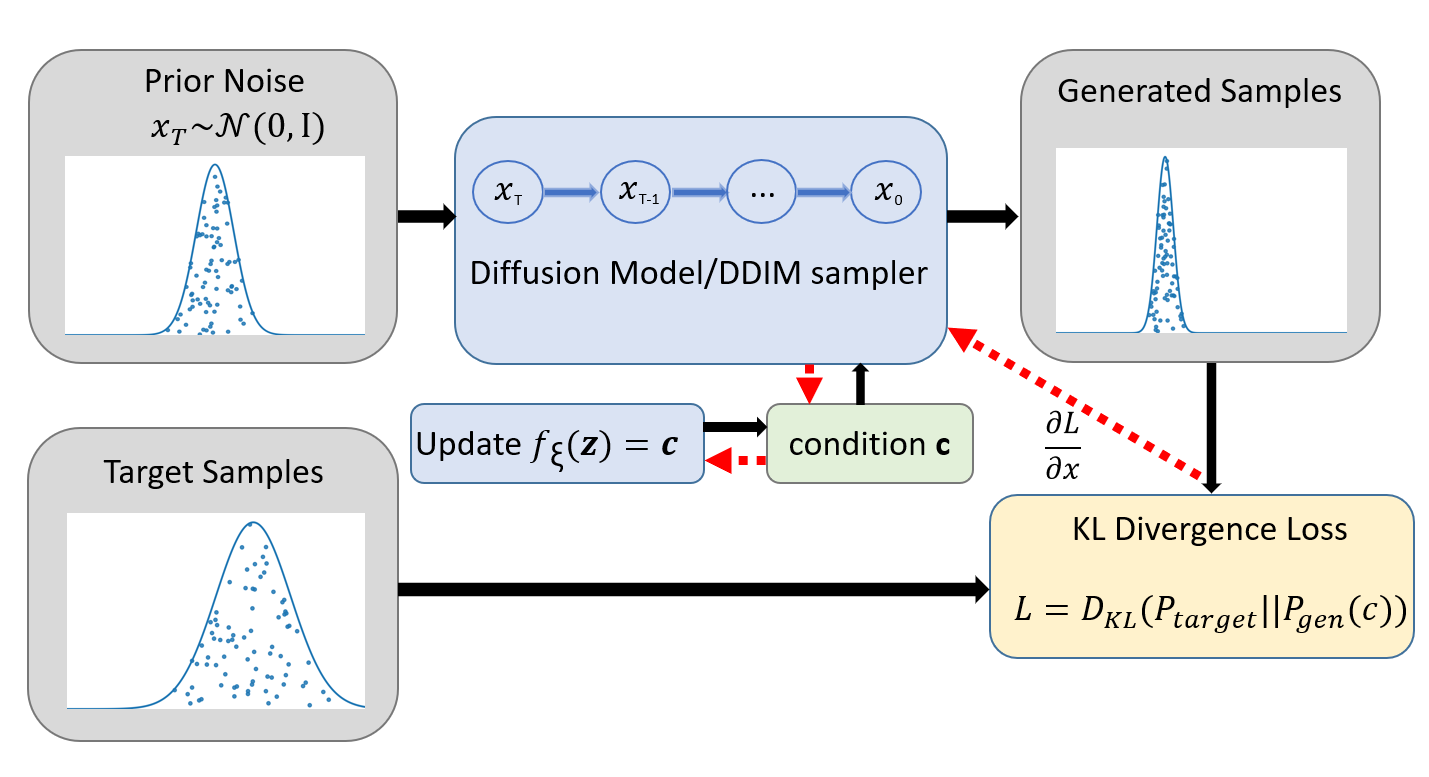}
    \caption{Schematic representation of the GB-FESO workflow. The diffusion model generates a large batch of samples with a given condition vector. The optimizer calculates the statistical loss between the generated and target ensembles and backpropagates gradients through the entire sampler path back to the condition, parameterized by an auxiliary neural network.}
    \label{fig:gb-feso}
\end{figure*}

\subsection{Gradient-based FES optimization}
\label{sec:gbfeso}

As noted above, after training, the conditional diffusion model is used as a differentiable
surrogate for the ensemble distribution. The model parameters are kept fixed,
and optimization is performed over the input condition rather than over the
generated samples themselves. In the context of FES design, this means that we
seek a condition $\boldsymbol{c}$ whose generated ensemble has the same probability
distribution, and therefore the same free-energy surface, as a prescribed
target ensemble.

The condition $\boldsymbol{c}$ represents the design variables that control the generated
distribution. These variables may be continuous, such as coefficients in a force-field parameterization, or discrete, such as bead types, chemical components, residue identities, or amino acid sequences. For a fixed condition, the diffusion sampler generates an ensemble of
configurations, which can be interpreted as samples from the distribution
\(\hat p_{\mathrm{DDPM}}(\boldsymbol{x}\mid \boldsymbol{c})\). The inverse-design task
is therefore to update $\boldsymbol{c}$ so that the generated distribution approaches the prescribed
target distribution \(\hat p_{\mathrm{target}}(\boldsymbol{x})\) (Fig.~\ref{fig:gb-feso}).

In the field of image and video generation, evaluation metrics are typically focused on the quality of individual samples, whereas the accuracy with which a
model reproduces the target distribution is addressed less often. For conformational ensembles, however, distributional accuracy is a primary objective. Thus, our goal is to match the ensemble distribution by backpropagating from a distribution-level loss, rather than from a comparison of individual samples.

This requires a quantitative, and ideally differentiable, metric that measures
agreement between the generated and target ensemble distributions. A natural
and widely used choice is the Kullback--Leibler (KL) divergence~\cite{KL}:
\begin{equation}
D_{\mathrm{KL}}(P \,\|\, Q)
=
\int p(\mathbf{x})\,\log\frac{p(\mathbf{x})}{q(\mathbf{x})}\,d\mathbf{x} .
\label{eq:kl1}
\end{equation}

Since the generated distribution has no explicit analytic form, the KL
divergence can be approximated using Gaussian kernel density estimation
(KDE)~\cite{KDE}, which yields a differentiable loss:
\begin{equation}
\begin{split}
D_{\mathrm{KL}}\!\left(
\hat p_{\mathrm{target}} \,\|\,
\hat p_{\mathrm{gen}}
\right)
&\approx
\frac{1}{N}\sum_{j=1}^{N}
\log
\frac{\hat p_{\mathrm{target}}(\mathbf{x}_j)}
{\hat p_{\mathrm{gen}}(\mathbf{x}_j)},\\
&\qquad \mathbf{x}_j \sim \hat p_{\mathrm{target}} .
\end{split}
\label{eq:kl2}
\end{equation}
Here
\begin{equation}
\begin{split}
\log p(\mathbf{x})
&\approx \log \hat p(\mathbf{x})\\
&= \log\!\left[
\frac{1}{N}\sum_{i=1}^{N}
\frac{1}{(2\pi)^{d/2}h^d}
\exp\!\left(
-\frac{\|\mathbf{x}-\mathbf{x}_i\|^2}{2h^2}
\right)
\right],
\end{split}
\label{eq:log_kde}
\end{equation}
and $h$ is the KDE bandwidth. Given an input condition $\mathbf{c}$, the final optimization objective can be written
as
\begin{equation}
\begin{split}
\mathbf{c}^\ast
= \arg\min_{\mathbf{c}\in\mathcal{C}}
\Bigl[ &
D_{\mathrm{KL}}\!\left(
\hat p_{\mathrm{target}}(\mathbf{x}) \,\|\,
\hat p_{\mathrm{DDPM}}(\mathbf{x}\mid \mathbf{c})
\right) \\
& + \lambda D_{\mathrm{KL}}\!\left(
\hat p_{\mathrm{DDPM}}(\mathbf{x}\mid \mathbf{c}) \,\|\,
\hat p_{\mathrm{target}}(\mathbf{x})
\right)
\Bigr],
\end{split}
\label{eq:kl3}
\end{equation}
where \(\mathcal{C}\) is the allowed domain of the design variables, and \(\lambda\) controls the contribution of the reverse KL term. In the simplest case, we set \(\lambda=0\).

As an alternative to the KL divergence, we can use the maximum mean discrepancy (MMD), a kernel-based metric for comparing probability distributions directly from samples. Unlike the KL divergence,
MMD does not require explicit density estimation and provides
a differentiable objective for distribution matching.
For target samples $\{\mathbf{x}_i\}_{i=1}^{N}$ and generated
samples $\{\mathbf{y}_j\}_{j=1}^{M}$, the squared MMD is
estimated as
\begin{equation}
\begin{split}
\mathrm{MMD}^2
={}& \frac{1}{N^2}\sum_{i,i'=1}^{N}
k(\mathbf{x}_i,\mathbf{x}_{i'})
+ \frac{1}{M^2}\sum_{j,j'=1}^{M}
k(\mathbf{y}_j,\mathbf{y}_{j'})\\
&- \frac{2}{NM}\sum_{i=1}^{N}
\sum_{j=1}^{M}k(\mathbf{x}_i,\mathbf{y}_j),
\end{split}
\label{eq:mmd}
\end{equation}
where $k$ is a positive-definite kernel, typically chosen
as a Gaussian kernel,
$k(\mathbf{x},\mathbf{y})
=\exp(-\|\mathbf{x}-\mathbf{y}\|^2/2\sigma^2)$,
with bandwidth $\sigma$.
For a characteristic kernel, the population MMD vanishes
if and only if the two distributions are identical.
Thus, MMD provides a good alternative distribution-level loss
that can be minimized~\cite{mmd1,mmd2}.

At each optimization step, while searching for the system design parameters that yield a desired FES, a synthetic ensemble is generated using the current condition $\mathbf{c}$, the statistical loss is evaluated against the target ensemble, and gradients are backpropagated through the sampling trajectory to update the condition while keeping the model parameters frozen.

To avoid being trapped in local minima during optimization, we use an overparameterization trick in which the condition $\mathbf{c}$ is generated by a small auxiliary neural network, $f_{\xi}(\mathbf{z})=\mathbf{c}$, where $\mathbf{z}$ is a trainable auxiliary vector. While direct optimization of the condition is possible, this trick significantly improves the method's performance.

A potential challenge in applying the proposed methodology with diffusion models is the need to backpropagate through dozens to hundreds of sampling steps. Nevertheless, previous work has demonstrated the feasibility of such optimization by differentiating directly through the sampling process~\cite{backprop1}, or by using approximate techniques such as adjoint ODE differentiation~\cite{backprop3} or gradient reparameterization~\cite{backprop4}. The requirement to compute gradients restricts the choice of sampler to differentiable formulations, thereby excluding certain popular stochastic samplers such as stochastic DDIM. In this work, we therefore employ a deterministic DDIM sampler, although other differentiable alternatives, including Euler- and Runge--Kutta-based solvers, could also be used~\cite{DDIM}.

Finally, we note that continuous conditioning variables can be optimized directly using gradient-based methods. Discrete conditioning variables, however, are not differentiable and, therefore, require a continuous relaxation during optimization. If the representation allows it, the force-field parameters can be optimized as continuous variables and then rounded to the closest matching label. Another possible approach is binary logit relaxation~\cite{louizos2018learning}; several alternative methods for handling discrete variables have also been proposed, including the Straight-Through Estimator (STE)~\cite{bengio2013estimating} and the Gumbel--Softmax relaxation~\cite{jang2017categorical,maddison2017concrete}.

\section{Results and discussion}
\label{sec:results}

\begin{figure}[tb]
    \centering
    \includefigure[width=\columnwidth]{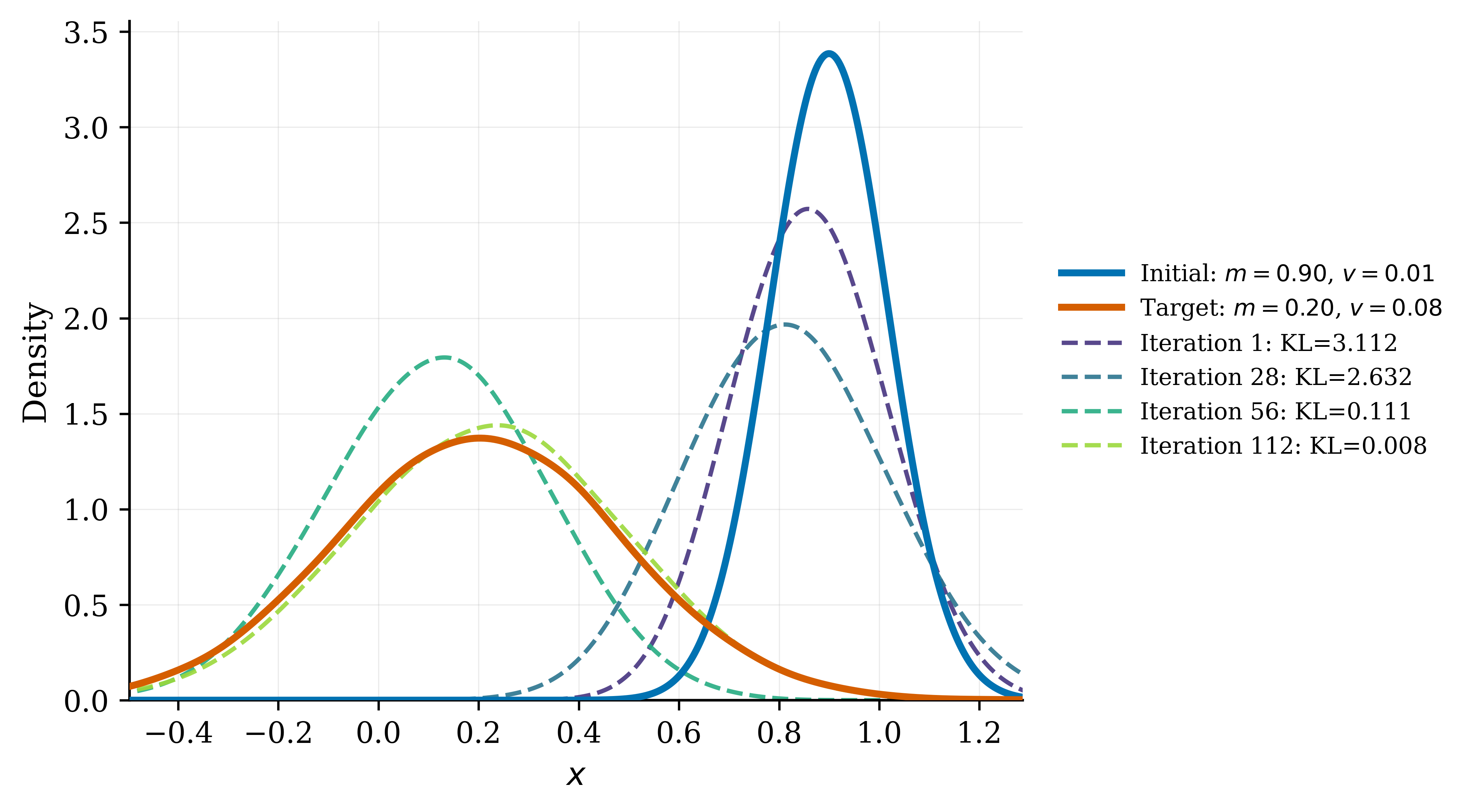}
    \caption{GB-FESO applied to a family of one-dimensional Gaussian distributions, where the generated samples are scalar values along the $x$-axis and the generative model's conditioning variables are the discretized mean and variance of the distribution from which the numbers are sampled.}
    \label{fig:gauss}
\end{figure}

\begin{figure*}[t]
    \centering
    \begin{minipage}[b]{0.45\textwidth}
        \centering
        \includefigure[width=\linewidth]{chain4.PNG}
        \panel{a}
    \end{minipage}
    \hfill
    \begin{minipage}[b]{0.53\textwidth}
        \centering
        \includefigure[width=\linewidth]{4bead_gener.PNG}
        \panel{b}
    \end{minipage}
    \caption{(a) Toy model described by LJ parameters $\varepsilon_1$ and $\varepsilon_2$ for nearest and next-nearest neighbors, respectively, and examples of possible configurations during the simulation. (b) Target vs. generated FES at the same LJ parameters: $\boldsymbol{\varepsilon}=(0.354; 0.346)$ and $\boldsymbol{\varepsilon}=(0.252; 0.835)$ with metastable states highlighted: A for folded, B for cross-folded and C for unfolded.}
    \label{fig:toy_overview}
\end{figure*}

\subsection{Applying GB-FESO to a domain of randomly generated Gaussian distributions}
\label{sec:gaussian}

As an initial proof of concept, we first investigate the proposed method in the context of a family of one-dimensional Gaussian distributions (Fig.~\ref{fig:gauss}). In this setting, the conditioning variables directly control the mean and variance of the generated distribution. Although intentionally simple, this example provides a well-controlled test case: the target distribution is known exactly, the conditioning variables have a clear interpretation, and the optimization outcome can be assessed directly.

The purpose of this experiment is to verify that the conditioning input of a trained diffusion model can be optimized to generate samples from a desired target distribution. As noted above, once the diffusion model has been trained, all model parameters are kept fixed. Starting from an initial condition associated with ensemble $I$, we seek a new condition whose generated ensemble matches a target ensemble $T$. Because sampling is performed using a deterministic DDIM procedure, the entire generation trajectory remains differentiable with respect to the relaxed conditioning variables, allowing gradients to propagate from the final loss [Eq.~\eqref{eq:kl3}] back to the condition parameters.

In the continuous version of the experiment, the mean and variance are represented directly as continuous conditioning variables. This provides the simplest possible setting and serves as a baseline for evaluating the optimization procedure. Successful optimization drives the generated distribution from the initial ensemble $I$ toward the target ensemble $T$ by appropriately adjusting the mean and variance.

To demonstrate that GB-FESO is also compatible with discrete conditioning variables, we encode the condition as a pair of integer indices, $(m_{\mathrm{idx}}, v_{\mathrm{idx}})$, corresponding to a discretized grid over the mean and variance. These integer values are represented as bit vectors and provided to the model as conditioning inputs. During optimization, the discrete bits are replaced by relaxed continuous variables, allowing gradients to propagate through the conditioning representation while retaining a discrete interpretation after decoding or rounding. For a given condition, samples are drawn as
\begin{equation}
x = \frac{m_{\mathrm{idx}}}{100}
+
\sqrt{\frac{v_{\mathrm{idx}}}{100}}\,\xi,
\qquad
\xi \sim \mathcal{N}(0,1).
\label{eq:gau}
\end{equation}
Here, \(m_{\mathrm{idx}}\) determines the Gaussian mean, \(v_{\mathrm{idx}}\) determines the variance, and \(\xi\) is a standard normally distributed random variable.

The DDPM is trained on a deliberately discontinuous set of conditions. The mean index is sampled from the integer ranges
\[
m_{\mathrm{idx}} \in [0,40] \cup [60,100],
\]
and the variance index from
\[
v_{\mathrm{idx}} \in [1,3] \cup [6,10].
\]
This split is used to test whether optimization of the condition vector can move the generated ensemble between distinct regions of the learned conditional distribution.

\begin{figure*}[t]
    \centering
    \includefigure[width=\textwidth]{traj_4_new.PNG}
    \caption{GB-FESO applied to the toy peptide. We start optimization from different initial $\boldsymbol{\varepsilon}$ values, optimize them to match a target distribution obtained from MD trajectories, and compare the result with an MD simulation at the optimized values (on the right). We show the initial generated FES and the intermediate steps leading to the optimal result.}
    \label{fig:optim}
\end{figure*}

\begin{figure*}[t]
    \centering
    \begin{minipage}[b]{0.49\textwidth}
        \centering
        \includefigure[width=\linewidth]{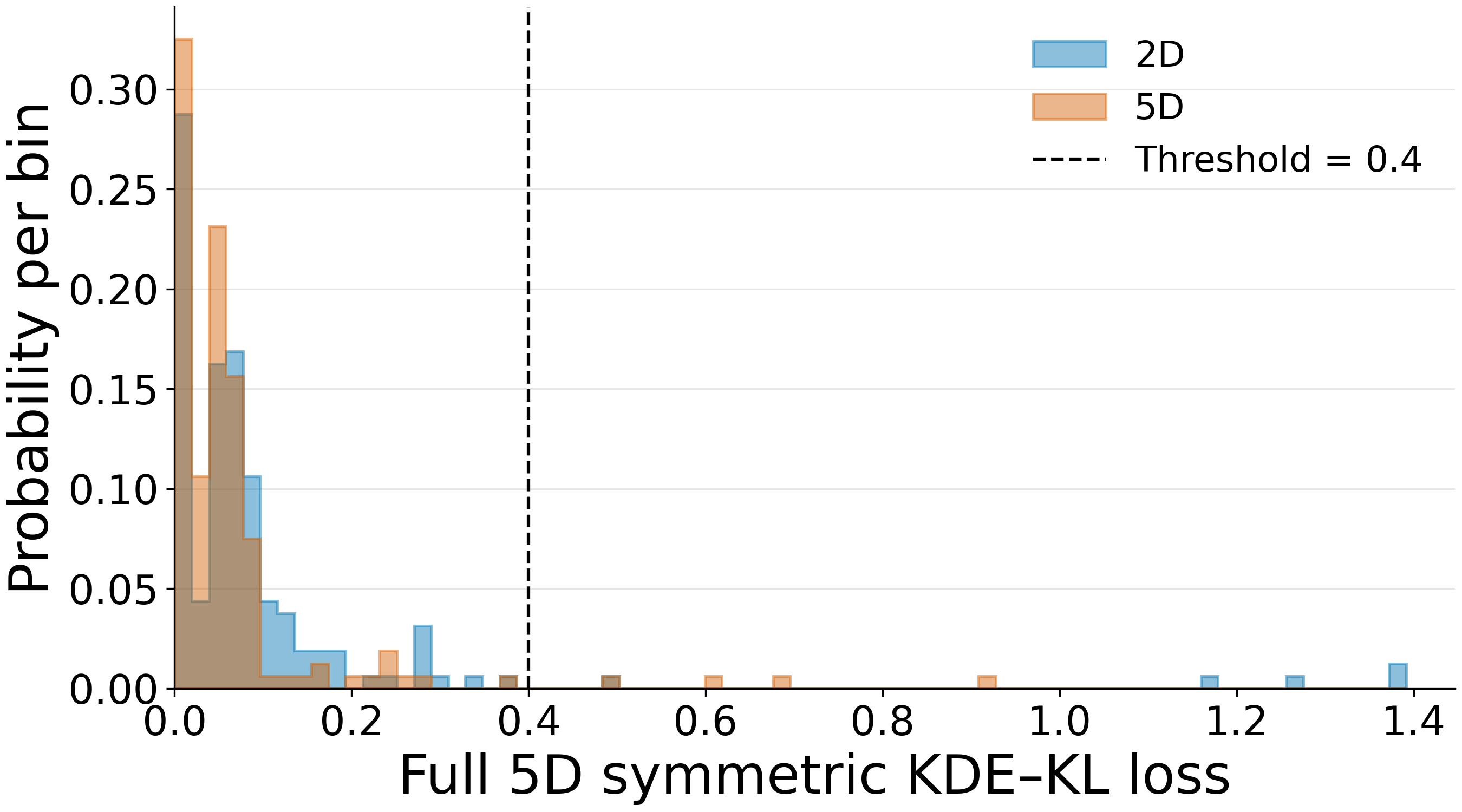}
        \panel{a}
    \end{minipage}
    \hfill
    \begin{minipage}[b]{0.49\textwidth}
        \centering
        \includefigure[width=\linewidth]{loss_histogram_mmd.PNG}
        \panel{b}
    \end{minipage}

    \vspace{0.5em}
    \begin{minipage}[b]{0.4\textwidth}
        \centering
        \includefigure[width=\linewidth]{violin.PNG}
        \panel{c}
    \end{minipage}
    \caption{(a) Distribution of the final KDE-KL loss (evaluated in full 5D) for runs optimized in 2D and in 5D. The dashed line marks the success threshold of 0.4. (b) Distribution of the final MMD loss with a threshold of 0.1. (c) Violin plots and medians of the errors in $\Delta G$ between the folded (A) and cross-folded (B) metastable states, in $k_BT$ units.}
    \label{fig:figmetrics}
\end{figure*}

This discrete formulation is important because many practical molecular or material design problems involve conditioning variables that are not naturally continuous. Examples include bead types, atom types, amino acid sequences, categorical labels, or other indexed descriptors. In such cases, directly optimizing a continuous physical parameter may not be meaningful, whereas optimizing a relaxed representation of a discrete condition provides a differentiable route toward selecting an appropriate category.

Our results show that, for this simple test case, GB-FESO is consistently able to reproduce the predefined target ensembles using the relaxed conditioning variables in both the continuous and discrete settings. The optimization trajectory shown in Fig.~\ref{fig:gauss} illustrates several intermediate ensembles generated throughout the optimization process. Notably, the procedure also demonstrates the ability to extrapolate beyond the training domain, with the optimized condition converging to mean and variance values that are not present in the training data while still generating the corresponding target distributions.

\subsection{4-bead toy peptide}
\label{sec:4bead}

In a second, more challenging case study, we investigate the ability to engineer the FES of a two-dimensional Lennard--Jones (LJ) four-particle toy peptide with five degrees of freedom and three metastable states [Fig.~\ref{fig:toy_overview}(a)]. Such a system is small enough that its equilibrium ensemble can be easily visualized, yet it still exhibits non-trivial conformational statistics due to the competition between its bonded constraints, non-bonded Lennard--Jones interactions~\cite{LennardJones} and entropy. The internal-coordinate representation of such a system is 5-dimensional and can be written as:
\[
\mathbf{x}
=
(r_{12},r_{23},r_{34},\theta_{123},\theta_{234}),
\]
where $r$ denotes distances and $\theta$ denotes bond angles. Ground-truth reference ensembles were generated using molecular dynamics simulations in LAMMPS~\cite{LAMMPS}. Non-bonded interactions are modeled with the LJ potential
\begin{equation}
V(r)=4\varepsilon\left[\left(\frac{\sigma}{r}\right)^{12}
-\left(\frac{\sigma}{r}\right)^6\right],
\label{eq:ljpot}
\end{equation}
where \(\varepsilon\) controls the interaction strength and \(\sigma\) sets the interaction
length scale. Interparticle bonds are modeled as harmonic,
\begin{equation}
U_{\mathrm{bond}}(r)=k(r-r_0)^2 ,
\label{eq:k}
\end{equation}
with a force constant \(k\) large enough to hold the chain together while still allowing substantial bond-length oscillations. The cutoff radii $R_{\mathrm{cut}}$ for the LJ potentials are set equal to 4.0.

\begin{figure*}[t]
    \centering
    \begin{minipage}[b]{0.4\textwidth}
        \centering
        \includefigure[width=\linewidth]{2ch_struck.PNG}
        \panel{a}
    \end{minipage}
    \hfill
    \begin{minipage}[b]{0.53\textwidth}
        \centering
        \includefigure[width=\linewidth]{2ch_gener.PNG}
        \panel{b}
    \end{minipage}
    \caption{(a) The three distinct metastable states observed in simulations of all pairs of interacting chains (from left to right: C, B and A). (b) Target vs. generated FES at the same LJ parameters, showing the basins of the metastable states. RMSD CVs are defined with respect to A and B.}
    \label{fig:2ch_overview}
\end{figure*}

\begin{figure*}[t]
    \centering
    \includefigure[width=\textwidth]{traj_2ch2.PNG}
    \caption{GB-FESO applied to the 2-chain system. From left to right: initial FES with random $\boldsymbol{\varepsilon}$;
    optimized generated FES; MD results for optimized $\boldsymbol{\varepsilon}$; generated FES with $\boldsymbol{\varepsilon}$ rounded to the nearest predefined particle-type values; target FES. Each plot lists the set of $\varepsilon$ values for the optimized chain; in all experiments, the second chain sequence remains fixed.}
    \label{fig:2chainopy}
\end{figure*}

\begin{figure*}[t]
    \centering
    \includefigure[width=0.88\textwidth]{mmd_violin.PNG}
    \caption{GB-FESO results for the 2-chain system. (a) Distribution of the final squared MMD loss, with an average of 0.00678. (b) Violin plot of the error between the MD and generated $\Delta G$ for states A and C, with an average of $0.135\,k_BT$.}
    \label{fig:2chresult}
\end{figure*}

If \(\sigma\) is fixed, the equilibrium distribution is controlled only by the LJ interaction strengths \(\varepsilon\). We therefore use \(\varepsilon\) as the conditioning variable for the generative model. For two interaction types, the condition can be written as
\(\boldsymbol{\varepsilon}=(\varepsilon_1,\varepsilon_2)\), corresponding to independently
controlled LJ pair coefficients, with $\varepsilon_1$ representing interactions between nearest neighbors and $\varepsilon_2$ representing interactions between next-nearest neighbors.

The four-bead toy peptide provides a convenient testbed for the optimization procedure. A fully trained DDPM with a linear noise schedule, a deterministic DDIM sampler, and a FiLM-conditioned MLP backbone~\cite{FiLM} is able to reproduce the correct distributions for both seen and unseen $\boldsymbol{\varepsilon}$ pairs [Fig.~\ref{fig:toy_overview}(b)].

Having established that the trained DDPM can generate conformational ensembles that are consistent with the ground truth, we move on to explore the proposed framework's ability to find the system design parameters that yield a desired ensemble-level FES. To better navigate the landscape, we use a two-step optimization procedure: first, a high learning rate with stochastic perturbations to escape local minima, followed by additional optimization steps with a lower learning rate. For evaluation, we are interested in finding a pair of $\boldsymbol{\varepsilon}$ values that produces an FES sufficiently similar to the desired one and thus yields a low KL/MMD loss.

First, we perform optimization in the full 5-dimensional space in a total of 160 runs, starting from low and high $\varepsilon_1$ and $\varepsilon_2$ regions, 40 runs for each starting region, using both KDE-KL and MMD losses. As noted above, the peptide allows three metastable states: folded, unfolded, and cross-folded [Fig.~\ref{fig:toy_overview}(a)]. Different combinations of $\boldsymbol{\varepsilon}$ lead to different populations of these states. We visually examined the optimized trajectories to check that the same metastable basins are populated in both the optimized and target FESs. We set the threshold for a successful optimization at $\sim$0.4 for the KL-based loss and $\sim$0.1 for the MMD-based loss, as these values reliably indicate that all correct basins are populated.

Our results show that the gradient-based optimization in the input space can adjust the LJ parameters such that the resulting distributions are sufficiently close to their target, as in the examples shown in Fig.~\ref{fig:optim}. As shown in figure~S1 in the supplementary material, the method can also be evaluated in the angle--distance representation. MMD-based optimization reached the desired threshold in all cases, while KDE-KL optimization had a success rate of 97.5\%, with 4 optimization trajectories ending unsuccessfully.
To further assess the accuracy of the optimized FESs, we compare the free-energy difference $\Delta G$ between the folded (A) and cross-folded (B) metastable states with the corresponding difference in the target FESs:
\[
    \frac{\Delta G}{k_BT}=\frac{G_B-G_A}{k_BT}=\ln\frac{P_A}{P_B},
\]
where $P_A$ and $P_B$ are the probabilities of states A and B, respectively. As shown in Fig.~\ref{fig:figmetrics}(c), optimization can achieve accurate $\Delta G$ predictions, deviating from the target on average by 0.159~$k_BT$ (KDE-KL) and 0.179~$k_BT$ (MMD).

Performing optimization in the full multidimensional space, including all the system's structural variables, is expected to be computationally demanding for larger systems with dozens or hundreds of degrees of freedom. To improve the scalability of the optimization procedure, it would therefore be advantageous to operate in a reduced representation of the system, using, for example, a set of collective variables. One such data-efficient approach, which we explored here, uses Harmonic Linear Discriminant Analysis (HLDA)~\cite{Dan2018,Piccini2018MetadynamicsDiscriminants,Mendels2018FoldingHLDA,MendelsParrinello2019}, which systematically constructs a low-dimensional set of collective variables that effectively separate the metastable states of the system.

Applying HLDA to a trajectory containing all three metastable states reveals (Table~\ref{tab:hlda}) that the peptide's bond angles play a more significant role than its interatomic distances in distinguishing between these states for the given trajectory. Accordingly, we performed GB-FESO using a KL-divergence loss computed only from the two bond-angle distributions, rather than from all structural degrees of freedom. As shown in Fig.~\ref{fig:figmetrics} and Table~\ref{tab:error}, the method is still able to accurately reproduce the target free-energy surface, with a minimal difference in the final optimization error, well below the desired threshold. For MMD, all runs remain successful, while KDE-KL success drops to 96.9\%. Additionally, we find that here too optimization can achieve accurate predictions of $\Delta G$, deviating from the target on average by 0.182~$k_BT$ (KDE-KL) and 0.174~$k_BT$ (MMD) [Fig.~\ref{fig:figmetrics}(c)].

\begin{table}[tb]
\caption{HLDA coefficients calculated from a single trajectory simulated with $\boldsymbol{\varepsilon}= (0.448; 0.287)$.}
\label{tab:hlda}
\begin{ruledtabular}
\begin{tabular}{lc}
Descriptor & Importance \\
\hline
$\theta_{123}$ & 0.2323 \\
$\theta_{234}$ & 0.2432 \\
$r_{12}$ & 0.1798 \\
$r_{23}$ & 0.1630 \\
$r_{34}$ & 0.1816 \\
\end{tabular}
\end{ruledtabular}
\end{table}

\begin{table}[tb]
\caption{Average loss and average error in the calculated free-energy difference $\Delta G$ between the folded (A) and cross-folded (B) metastable states, computed from FESs in which both states are present.}
\label{tab:error}
\begin{ruledtabular}
\begin{tabular}{lccc}
Run & Loss & $\Delta G$, $k_BT$ & Loss below threshold \\
\hline
5D, KDE-KL & 0.0674 & 0.159 & 97.5\% \\
5D, MMD & 0.0071 & 0.179 & 100\% \\
2D, KDE-KL & 0.1013 & 0.182 & 96.9\% \\
2D, MMD & 0.0082 & 0.174 & 100\% \\
\end{tabular}
\end{ruledtabular}
\end{table}

\subsection{2-chain system}
\label{sec:2chain}

To examine the method's capabilities in a larger optimization space, we consider a two-dimensional system consisting of two interacting 5-bead chains with varying sequences of particle types. This system can serve as an initial simplified model for the prediction and inverse design of interacting peptide or disordered protein segments, a problem of significant scientific and practical importance~\cite{Krueger2026IDPDesign}. As in the previous example, the particles interact through Lennard--Jones forces. However, the bonds are defined to be stiffer to keep the distances between adjacent beads approximately fixed.

We now represent the system using a distance matrix containing the pairwise distances between all beads. For each bead, $\varepsilon$ is selected from a set of eight values ranging from 0.25 to 1.08125, with interactions between different bead types calculated in LAMMPS with the geometric mixing rule. When the chains are in contact, the system can occupy one of three metastable states [Fig.~\ref{fig:2ch_overview}(a)]: ``Extended'' (A), ``Interlocked'' (B) and ``Wrapped'' (C). Frames in which the chains are not in contact are discarded. At least two of these states are populated in all simulations, but their relative populations vary significantly depending on the chain sequences.

We randomly initialize the $\varepsilon$ values for all 10 beads. However, during optimization, we keep the sequence of one chain fixed and optimize only that of the second chain, $\boldsymbol{\varepsilon} = [\varepsilon_1,\varepsilon_2,\varepsilon_3,\varepsilon_4,\varepsilon_5]$. As in the 4-bead optimization, to reduce computational cost and obtain an easily interpretable representation, we perform the optimization in a reduced-dimensional space. Here, we use two CVs, defined as the distance-based root-mean-square deviations (RMSDs) from reference frames corresponding to configurations A and B [Fig.~\ref{fig:2ch_overview}(b)].

For optimization, we use the MMD loss, which performed better for the 4-bead system. To make the optimization semi-discrete, we round the optimized conditions to the nearest predefined particle-type values. As illustrated in Fig.~\ref{fig:2chainopy}, the FESs obtained from the rounded and unrounded optimized conditions were very similar across all inspected runs. We apply GB-FESO to 120 randomly initialized chain pairs, optimizing their sequences to better reproduce the target FESs. The optimizer achieves very low loss values, with an average MMD loss of 0.00678 across all runs and only eight outliers with $\mathrm{MMD} \geq 0.02$ [Fig.~\ref{fig:2chresult}(a)]. The average error in the free-energy difference $\Delta G$ between metastable states A and C is $0.135\,k_BT$ [Fig.~\ref{fig:2chresult}(b)], comparable to that obtained for the 4-bead system.

\section{Conclusions}
\label{sec:conclusions}

In this work, we have proposed GB-FESO, a framework for the inverse design of systems with target free-energy landscapes using conditional diffusion models as differentiable ensemble generators. Rather than optimizing individual conformations or selecting samples that satisfy a target property, our approach directly optimizes the conditioning variables that determine the generated ensemble. This formulation enables direct comparison between the generated and target distributions through an ensemble-level loss, while updating the design variables by backpropagating gradients through the deterministic diffusion sampling trajectory.

We first demonstrate the method on ensembles drawn from simple one-dimensional Gaussian distributions, where the conditioning variables correspond directly to the distribution parameters. This controlled setting confirms that the conditioning variables of a trained and frozen generative model can be optimized to transform one ensemble distribution into another. We then apply the same principle to a toy peptide model governed by Lennard--Jones interaction parameters. In this more physically motivated setting, the method successfully optimizes the interaction strengths, starting from previously unseen initial values, to reproduce target free-energy surfaces. Finally, we scale up the system from a single 4-bead chain to a pair of interacting 5-bead chains. Even in this larger optimization space, we find that GB-FESO successfully optimizes the system's ensemble behavior through sequence optimization.

These results suggest that conditional diffusion models provide a practical route for ensemble-level molecular and materials design, particularly in settings where the desired outcome is naturally expressed as a distribution rather than as a single optimized structure. At the same time, several challenges must be addressed before this approach can be extended to larger and more realistic molecular systems. For very large neural networks and/or high-dimensional free-energy landscapes, straightforward backpropagation through the diffusion sampling trajectory may become computationally inefficient or unstable. Moreover, the KDE-based approximation of the KL divergence objective can give rise to a rough optimization landscape, hindering convergence. As shown in this work, however, using the overparameterization trick and an alternative statistical loss, such as MMD, can substantially mitigate these issues. Furthermore, improving scalability and robustness in higher-dimensional settings may require alternative distributional objectives, such as Wasserstein-based metrics~\cite{Wasserstein}.

Overall, the proposed framework represents a promising first step toward differentiable free-energy-surface design. Although further methodological developments are needed, the results demonstrate that optimizing the conditioning variables of a frozen generative model can effectively steer generated ensembles toward target thermodynamic behavior and target free-energy landscapes.

\begin{acknowledgments}
The authors acknowledge support from the Israel Science Foundation (ISF) under grant number 1181/24. The authors thank Dr. Assaf Shocher for a helpful discussion.

AI disclosure: During the preparation of this work, the authors used Claude.ai and ChatGPT to proofread the manuscript, format the LaTeX template, assist with writing and debugging code, and search for and verify references. The authors reviewed, validated and critically assessed all AI-assisted outputs and take full responsibility for the content of the publication.

\medskip
E.Z. and D.M. conceptualized and planned the study, analyzed the data, and wrote the manuscript. E.Z. implemented the methods, performed the simulations, performed the computational analyses, and generated the results. D.M. acquired funding and supervised the project.

\medskip
The authors declare no competing interests.
\end{acknowledgments}

\section*{Data availability statement}

The code, model weights and the dataset scripts used to generate and analyze the data supporting this article are openly available in Zenodo at \url{https://doi.org/10.5281/zenodo.24032550}, accession number: 10.5281/zenodo.24032550. The development version of the code is available on GitHub at \url{https://github.com/MendelsResearchGroup/Gradient-Based-Free-Energy-Surface-Optimization/tree/dev-latest} and is released under the MIT License.

\section*{Supplementary data}

Additional distance--angle free-energy-surface examples after optimization are provided in the supplementary material.

\appendix

\section{Gaussian test case}
\label{app:gaussian}

For the one-dimensional proof of concept, the training data are generated dynamically during training from Gaussian distributions whose mean and variance are controlled by two integer-valued condition indices,
\[
m_{\mathrm{idx}}, \; v_{\mathrm{idx}} .
\]

The Gaussian DDPM uses \(100\) diffusion steps with a linear noise schedule from
\(\beta_1=10^{-4}\) to \(\beta_T=0.1\). The model is trained for \(6000\) optimization steps with a batch size of \(1024\), using the Adam~\cite{adam} optimizer with a learning rate of \(2\times 10^{-4}\). The model is trained with the standard DDPM noise-prediction mean-squared-error loss per Eq.~\eqref{eq:mse}.

In the inverse-optimization example shown in Fig.~\ref{fig:gauss}, the initial and target distributions are
\[
I: (m_{\mathrm{idx}},v_{\mathrm{idx}})=(90,1),
\qquad
T: (m_{\mathrm{idx}},v_{\mathrm{idx}})=(20,8).
\]
The discrete indices are represented by binary encodings: seven bits for the mean index and four bits for the variance index. During optimization, these binary variables are replaced by relaxed continuous logits, allowing gradients to propagate through the deterministic DDIM sampler. The relaxed bits are evaluated using a sigmoid--Gumbel relaxation with a temperature \(\tau=0.7\).

At each optimization step, \(2048\) samples are generated using a deterministic DDIM sampler with \(40\) sampling steps. The condition logits are optimized for 140 iterations using Adam with a learning rate of \(0.12\). The loss is the KDE-based KL divergence between the generated distribution and the target distribution, evaluated on a one-dimensional grid over \(x\in[-1,2]\) with a Gaussian KDE bandwidth of \(h=0.06\). After optimization, the relaxed condition is decoded back to integer indices, while the generated density is compared visually against the initial and target Gaussian distributions.

\section{Molecular dynamics simulations}
\label{app:md}

\subsection{Four-bead chain}
\label{app:md-four-bead}

The reference ensembles for the first toy peptide were generated with LAMMPS using a two-dimensional four-bead chain. Each bead had a distinct particle type, and consecutive beads were connected by harmonic bonds to keep the chain intact while still allowing bond-length fluctuations. Non-bonded interactions were described by a Lennard--Jones potential with fixed \(\sigma=1\) and $R_{\mathrm{cut}}=4.0$. Same-type interactions were set to zero, while cross-type interactions defined the tunable energy landscape.

For each trajectory, two independent LJ energy parameters were randomly selected in the interval
\[
\varepsilon_1,\varepsilon_2 \in [0.21,1.55].
\]
The first parameter, \(\varepsilon_1\), was assigned to bonded-neighbor pairs,
\[
(1,2),\ (2,3),\ (3,4),
\]
while the second parameter, \(\varepsilon_2\), was assigned to next-nearest-neighbor pairs,
\[
(1,3),\ (2,4).
\]
The remaining long-range pair \((1,4)\) was kept fixed at $\varepsilon_3=0.5$. Each LAMMPS input file therefore defines one trajectory and one condition vector,
\[
\boldsymbol{\varepsilon}=(\varepsilon_1,\varepsilon_2).
\]

The simulations were performed in reduced Lennard--Jones units with Langevin dynamics at a low temperature of $T=0.09$. The resulting dump files were postprocessed into internal coordinates rather than Cartesian coordinates. For every saved frame, the sample vector was
\[
\mathbf{x}
=
(r_{12},r_{23},r_{34},\theta_{123},\theta_{234}),
\]
where the three \(r\) values are adjacent bond distances and the two \(\theta\) values are signed bond angles. These five-dimensional internal-coordinate samples were used as the training data for the conditional diffusion model. The corresponding \((\varepsilon_1,\varepsilon_2)\) values were extracted directly from the LAMMPS pair coefficients and used as the conditioning variables.

\subsection{Two-chain system}
\label{app:md-two-chain}

Likewise, the two-dimensional two-chain toy peptide was simulated in LAMMPS as ten beads (five in each chain), each of a separate type, with $\varepsilon$ selected randomly from eight values between 0.25 and 1.08125, making the parameters semi-discrete. The bond coefficient was set to 80, making the chains rigid, and the LJ cutoff radius was set to 4.0. Simulations were conducted at a temperature of 0.13, and frames in which the chains were disconnected (above the 2.2 threshold) were discarded. This time, the system was represented as a distance matrix between all beads in the system. This accounts for nine degrees of freedom and ten optimization parameters in total, $\boldsymbol{\varepsilon}=(\varepsilon_1,\varepsilon_2, \ldots , \varepsilon_{10})$. However, we optimize only the first chain, i.e., five parameters in total.

\section{Network architecture, training and inverse optimization}
\label{app:training}

\begin{figure*}[t]
    \centering
    \includefigure[width=0.83\textwidth]{bw_sens.PNG}
    \caption{GB-FESO performance with various bandwidth values for the KDE approximation and MMD kernel. For a correct comparison, all results are evaluated with the same bandwidth.}
    \label{fig:sens}
\end{figure*}

\begin{table*}[t]
\caption{Performance comparison against random search.}
\label{tab:compare}
\begin{ruledtabular}
\begin{tabular}{lccc}
Method & Included Runs & Average execution time (s) & Hardware \\
\hline
GB-FESO, 5D & 160 & 88.9 (MMD)/70.4 (KDE-KL) & GPU NVIDIA RTX 5090 \\
GB-FESO, 2D & 160 & 79.4 (MMD)/62.5 (KDE-KL) & GPU NVIDIA RTX 5090 \\
Random search (4 beads) & 160 & 1966.4 & CPU AMD Ryzen 7 7700 \\
Bayesian optimization (4 beads) & 160 & 25.6 & CPU AMD Ryzen 7 7700 \\
\hline
GB-FESO (2 chains) & 40 & 466.5 & GPU NVIDIA RTX 5090 \\
Bayesian optimization (2 chains) & 29 & 1942 & CPU AMD Ryzen 7 7700 \\
\end{tabular}
\end{ruledtabular}
\end{table*}

For both systems, the model learns a
conditional ensemble distribution
\[
p_{\theta}(\mathbf{x}\mid \boldsymbol{\varepsilon}),
\]
where changes in the condition correspond to changes in the interaction
parameters that define the system.

The 4-bead denoising network is a FiLM-conditioned multilayer perceptron built from residual blocks, each containing layer normalization, FiLM scale-and-shift modulation, two linear layers, a SiLU activation and dropout with probability 0.05. The diffusion timestep is encoded with a 64-dimensional sinusoidal embedding, and the timestep and condition embeddings are mapped through 128-dimensional conditioning layers. At each diffusion step, the network receives the noisy configuration vector, the timestep and the condition vector, and it is trained to predict the Gaussian noise added in the forward process using the standard DDPM mean-squared-error objective [Eq.~\eqref{eq:mse}]. For the 4-bead system, the network has a hidden width of 256 and seven residual blocks, and the condition is always provided as an input. For the 2-chain system, the network is a convolutional network acting on the $10\times10$ distance matrix, with 64 channels and six FiLM-conditioned residual blocks, each containing group normalization and two convolutional layers, with dilations of 1, 1, 2, 2, 1 and 1 across the blocks. This model is trained with classifier-free guidance, with the condition dropped with probability 0.1. We also use chain-reversal and chain-swap augmentation to account for the symmetry of the system.

Both models use a linear noise schedule with 1000 diffusion steps and are trained for 80 epochs with batch size 256 using AdamW with weight decay \(10^{-4}\) (\(10^{-2}\) for the 2-chain system). The learning rate is \(1.5\times10^{-4}\) for the 4-bead system and \(2\times10^{-4}\) for the 2-chain system. The training data consist of 400 trajectories of 20\,000 shuffled frames each for the 4-bead system, and 4846 trajectories of 2048--5000 frames each for the 2-chain system.

For inverse FES optimization, the trained DDPM parameters are frozen and only
the input condition is updated. At each optimization step, an ensemble is generated from the
current condition using the deterministic DDIM sampler. For the 4-bead case, the generated and target
ensembles are compared in the full internal-coordinate space (5D), so that both bond-length fluctuations and angular conformational changes contribute to the optimization objective, and, in a second set of runs, in the reduced angle space (2D). The loss is computed as the KDE-based KL divergence or MMD between the generated and target distributions, per Eqs.~\eqref{eq:kl3} and \eqref{eq:mmd}. Our calculations have shown that excluding the reverse term does not significantly affect performance; we therefore set $\lambda = 0$ to reduce the computational cost.
For the 2-chain system, we optimize within the space spanned by two RMSD CVs [defined with respect to metastable states A and B, respectively; Fig.~\ref{fig:2ch_overview}(b)], using the same statistical loss.

Optimization is performed in two stages. In the first stage, the condition is optimized with a
relatively large learning rate while stochastic perturbations are added to help
escape local minima. In the second stage, if the loss remains above a chosen threshold (0.1 for both KDE-KL and MMD), the condition is further refined with a lower learning rate and without additional perturbations. This procedure is
used to improve convergence toward a condition whose generated ensemble matches the target distribution. For this case, we use a KDE bandwidth of $h=0.5$ and an MMD bandwidth of $\sigma=0.32$, calculated from 1024--4096 samples per optimization step. Bandwidth values were chosen based on a sensitivity analysis (Fig.~\ref{fig:sens}). In general, however, a single bandwidth is not required, and multi-bandwidth optimization can in principle further smooth the loss.

The purpose of the optimization is not necessarily to recover the exact target
values of $\boldsymbol{\varepsilon}$. Since, in the broader case, different interaction parameters might produce similar ensemble statistics, success is instead defined
by the agreement between the generated and target distributions. Thus, the
optimized condition is evaluated according to the final statistical loss between the
generated and target internal-coordinate ensembles.

\section{Computational details}
\label{app:computational}

A central advantage of GB-FESO over derivative-free optimization is the way its cost scales with the number of design variables. For a fixed conditional network, ensemble size and number of diffusion steps, the cost of a single GB-FESO iteration is dominated by one forward and one backward pass through the sampling trajectory. Reverse-mode differentiation yields the full gradient with respect to all $N$ conditioning variables at a cost comparable to a single forward evaluation, so the per-iteration cost is essentially independent of $N$, and the total cost grows only through the number of iterations required for convergence. Timings of GPU- and CPU-based procedures are difficult to compare directly because they depend strongly on hardware. With this caveat, Table~\ref{tab:compare} compares GB-FESO with Bayesian optimization using the tree-structured Parzen estimator (TPE)~\cite{bergstra2011,watanabe2023}, as implemented in the Optuna library~\cite{optuna}, with both methods run to the same loss threshold. GB-FESO is slower for the 4-bead system, which has two design variables, but about four times faster for the 2-chain system, which has five. Both methods are substantially faster than random search.



\begin{thebibliography}{99}

\bibitem{Butler2018MLMolecularMaterials}
K. T. Butler, D. W. Davies, H. Cartwright, O. Isayev, and A. Walsh, Machine learning for molecular and materials science, \doilink{10.1038/s41586-018-0337-2}{Nature \textbf{559}, 547 (2018)}.

\bibitem{SanchezLengeling2018InverseMolecularDesign}
B. Sanchez-Lengeling and A. Aspuru-Guzik, Inverse molecular design using machine learning: generative models for matter engineering, \doilink{10.1126/science.aat2663}{Science \textbf{361}, 360 (2018)}.

\bibitem{Rives2021ProteinLM}
A. Rives \textit{et al.}, Biological structure and function emerge from scaling unsupervised learning to 250 million protein sequences, \doilink{10.1073/pnas.2016239118}{Proc. Natl. Acad. Sci. USA \textbf{118}, e2016239118 (2021)}.

\bibitem{Shteingolts2026StructureOnly}
S. A. Shteingolts, S. N. Salman, and D. Mendels, Enabling structure-only initialization and out-of-distribution generalization in GNN-based molecular dynamics simulators, \doilink{10.1021/acs.jctc.6c01145}{J. Chem. Theory Comput. (to be published)}.

\bibitem{SanchezGonzalez2020GraphNetworks}
A. Sanchez-Gonzalez, J. Godwin, T. Pfaff, R. Ying, J. Leskovec, and P. Battaglia, Learning to simulate complex physics with graph networks, in \href{https://proceedings.mlr.press/v119/sanchez-gonzalez20a.html}{\textit{Proceedings of the 37th International Conference on Machine Learning}}, PMLR Vol.~119 (2020), pp.~8459--8468.

\bibitem{diffusionOG}
J. Ho, A. Jain, and P. Abbeel, Denoising diffusion probabilistic models, in \href{https://proceedings.neurips.cc/paper/2020/file/4c5bcfec8584af0d967f1ab10179ca4b-Paper.pdf}{\textit{Advances in Neural Information Processing Systems}}, Vol.~33 (2020), pp.~6840--6851.

\bibitem{AlphaFold}
J. Jumper \textit{et al.}, Highly accurate protein structure prediction with AlphaFold, \doilink{10.1038/s41586-021-03819-2}{Nature \textbf{596}, 583 (2021)}.

\bibitem{BioEmu}
S. Lewis \textit{et al.}, Scalable emulation of protein equilibrium ensembles with generative deep learning, \doilink{10.1126/science.adv9817}{Science \textbf{389}, eadv9817 (2025)}.

\bibitem{RFdiffusion}
J. L. Watson \textit{et al.}, De novo design of protein structure and function with RFdiffusion, \doilink{10.1038/s41586-023-06415-8}{Nature \textbf{620}, 1089 (2023)}.

\bibitem{Chroma}
J. B. Ingraham \textit{et al.}, Illuminating protein space with a programmable generative model, \doilink{10.1038/s41586-023-06728-8}{Nature \textbf{623}, 1070 (2023)}.

\bibitem{FoldingDiff}
K. E. Wu, K. K. Yang, R. van den Berg, S. Alamdari, J. Y. Zou, A. X. Lu, and A. P. Amini, Protein structure generation via folding diffusion, \doilink{10.1038/s41467-024-45051-2}{Nat. Commun. \textbf{15}, 1059 (2024)}.

\bibitem{FrameDiff}
J. Yim, B. L. Trippe, V. De Bortoli, E. Mathieu, A. Doucet, R. Barzilay, and T. Jaakkola, SE(3) diffusion model with application to protein backbone generation, in \href{https://proceedings.mlr.press/v202/yim23a.html}{\textit{Proceedings of the 40th International Conference on Machine Learning}}, PMLR Vol.~202 (2023), pp.~40001--40039.

\bibitem{ExEnDiff}
Y. Liu, Z. Yu, R. J. Lindsay, G. Lin, M. Chen, A. Sahoo, and S. M. Hanson, ExEnDiff: an experiment-guided diffusion model for protein conformational ensemble generation, \doilink{10.1103/PRXLife.3.023013}{PRX Life \textbf{3}, 023013 (2025)}.

\bibitem{MinimumExcessWorkGuidance}
C. Kara, T. H\"oppe, E. Angelis, J. M. Schreiner, S. Bauer, A. Dittadi, and S. Olsson, Minimum-excess-work guidance: score-based sampling with experimental data or sparse restraints, \doilink{10.1021/acs.jctc.6c00080}{J. Chem. Theory Comput. \textbf{22}, 5838 (2026)}.

\bibitem{STARLING}
B. Novak, J. M. Lotthammer, R. J. Emenecker, and A. S. Holehouse, Accurate predictions of disordered protein ensembles with STARLING, \doilink{10.1038/s41586-026-10141-2}{Nature \textbf{652}, 240 (2026)}.

\bibitem{Stenstrom2024LigandRecognition}
O. Stenstr\"om, C. Diehl, K. Modig, and M. Akke, Ligand-induced protein transition state stabilization switches the binding pathway from conformational selection to induced fit, \doilink{10.1073/pnas.2317747121}{Proc. Natl. Acad. Sci. USA \textbf{121}, e2317747121 (2024)}.

\bibitem{Spoendlin2025AntibodyFlexibility}
F. C. Spoendlin \textit{et al.}, Predicting the conformational flexibility of antibody and T cell receptor complementarity-determining regions, \doilink{10.1038/s42256-025-01131-6}{Nat. Mach. Intell. \textbf{7}, 1755 (2025)}.

\bibitem{Nussinov2026Allostery}
R. Nussinov, C. Regev, and H. Jang, Leveraging conformational ensembles in allosteric drug discovery, \doilink{10.1016/j.tips.2026.01.006}{Trends Pharmacol. Sci. \textbf{47}, 276 (2026)}.

\bibitem{Frauenfelder1991}
H. Frauenfelder, S. G. Sligar, and P. G. Wolynes, The energy landscapes and motions of proteins, \doilink{10.1126/science.1749933}{Science \textbf{254}, 1598 (1991)}.

\bibitem{TorrieValleau1977}
G. M. Torrie and J. P. Valleau, Nonphysical sampling distributions in Monte Carlo free-energy estimation: umbrella sampling, \doilink{10.1016/0021-9991(77)90121-8}{J. Comput. Phys. \textbf{23}, 187 (1977)}.

\bibitem{LaioParrinello2002}
A. Laio and M. Parrinello, Escaping free-energy minima, \doilink{10.1073/pnas.202427399}{Proc. Natl. Acad. Sci. USA \textbf{99}, 12562 (2002)}.

\bibitem{Henin2022}
J. H\'enin, T. Leli\`evre, M. R. Shirts, O. Valsson, and L. Delemotte, Enhanced sampling methods for molecular dynamics simulations, \doilink{10.33011/livecoms.4.1.1583}{Living J. Comput. Mol. Sci. \textbf{4}, 1583 (2022)}.


\bibitem{Medaparambath2026}
M. Medaparambath, A. Zhilkin, and D. Mendels, Collective variable-guided engineering of the free-energy surface of a small peptide, \doilink{10.1021/acs.jcim.6c00674}{J. Chem. Inf. Model. \textbf{66}, 8465 (2026)}.

\bibitem{Alex2025}
A. Zhilkin, M. Medaparambath, and D. Mendels, Guiding peptide kinetics via collective-variable tuning of free-energy barriers, \doilink{10.1021/acs.jctc.6c00418}{J. Chem. Theory Comput. \textbf{22}, 4573 (2026)}.

\bibitem{AbramsBussi2014}
C. Abrams and G. Bussi, Enhanced sampling in molecular dynamics using metadynamics, replica-exchange, and temperature-acceleration, \doilink{10.3390/e16010163}{Entropy \textbf{16}, 163 (2014)}.

\bibitem{Dan2022}
D. Mendels and J. J. de Pablo, Collective variables for free energy surface tailoring: understanding and modifying functionality in systems dominated by rare events, \doilink{10.1021/acs.jpclett.2c00317}{J. Phys. Chem. Lett. \textbf{13}, 2830 (2022)}.

\bibitem{Dan2023}
D. Mendels, F. Byl\'ehn, T. W. Sirk, and J. J. de Pablo, Systematic modification of functionality in disordered elastic networks through free energy surface tailoring, \doilink{10.1126/sciadv.adf7541}{Sci. Adv. \textbf{9}, eadf7541 (2023)}.

\bibitem{SD}
R. Rombach, A. Blattmann, D. Lorenz, P. Esser, and B. Ommer, High-resolution image synthesis with latent diffusion models, in \textit{Proceedings of the IEEE/CVF Conference on Computer Vision and Pattern Recognition (CVPR)} (2022), pp.~10684--10695.

\bibitem{KL}
S. Kullback and R. A. Leibler, On information and sufficiency, \doilink{10.1214/aoms/1177729694}{Ann. Math. Stat. \textbf{22}, 79 (1951)}.

\bibitem{KDE}
E. Parzen, On estimation of a probability density function and mode, Ann. Math. Stat. \textbf{33}, 1065 (1962).

\bibitem{mmd1}
A. Gretton, K. M. Borgwardt, M. J. Rasch, B. Sch\"olkopf, and A. Smola, A kernel two-sample test, \href{https://www.jmlr.org/papers/v13/gretton12a.html}{J. Mach. Learn. Res. \textbf{13}, 723 (2012)}.

\bibitem{mmd2}
A. M\"uller, Integral probability metrics and their generating classes of functions, \doilink{10.2307/1428011}{Adv. Appl. Probab. \textbf{29}, 429 (1997)}.

\bibitem{backprop1}
B. Wallace, A. Gokul, S. Ermon, and N. Naik, End-to-end diffusion latent optimization improves classifier guidance, in \textit{Proceedings of the IEEE/CVF International Conference on Computer Vision (ICCV)} (2023), pp.~7280--7290, \href{https://arxiv.org/abs/2303.13703}{arXiv:2303.13703}.

\bibitem{backprop3}
J. Pan, H. Yan, J. H. Liew, J. Feng, and V. Y. F. Tan, Towards accurate guided diffusion sampling through symplectic adjoint method, \href{https://arxiv.org/abs/2312.12030}{arXiv:2312.12030}.

\bibitem{backprop4}
D. Watson, W. Chan, J. Ho, and M. Norouzi, Learning fast samplers for diffusion models by differentiating through sample quality, in \href{https://openreview.net/forum?id=VFBjuF8HEp}{\textit{International Conference on Learning Representations (ICLR)}} (2022).

\bibitem{DDIM}
J. Song, C. Meng, and S. Ermon, Denoising diffusion implicit models, in \textit{International Conference on Learning Representations (ICLR)} (2021), \href{https://arxiv.org/abs/2010.02502}{arXiv:2010.02502}.

\bibitem{louizos2018learning}
C. Louizos, M. Welling, and D. P. Kingma, Learning sparse neural networks through $L_0$ regularization, in \textit{International Conference on Learning Representations (ICLR)} (2018), \href{https://arxiv.org/abs/1712.01312}{arXiv:1712.01312}.

\bibitem{bengio2013estimating}
Y. Bengio, N. L\'eonard, and A. Courville, Estimating or propagating gradients through stochastic neurons for conditional computation, \href{https://arxiv.org/abs/1308.3432}{arXiv:1308.3432}.

\bibitem{jang2017categorical}
E. Jang, S. Gu, and B. Poole, Categorical reparameterization with Gumbel-Softmax, in \textit{International Conference on Learning Representations (ICLR)} (2017), \href{https://arxiv.org/abs/1611.01144}{arXiv:1611.01144}.

\bibitem{maddison2017concrete}
C. J. Maddison, A. Mnih, and Y. W. Teh, The concrete distribution: a continuous relaxation of discrete random variables, in \textit{International Conference on Learning Representations (ICLR)} (2017), \href{https://arxiv.org/abs/1611.00712}{arXiv:1611.00712}.

\bibitem{LennardJones}
J. E. Jones, On the determination of molecular fields. I. From the variation of the viscosity of a gas with temperature, Proc. R. Soc. A \textbf{106}, 441 (1924).

\bibitem{LAMMPS}
S. Plimpton, Fast parallel algorithms for short-range molecular dynamics, J. Comput. Phys. \textbf{117}, 1 (1995).

\bibitem{FiLM}
E. Perez, F. Strub, H. de Vries, V. Dumoulin, and A. Courville, FiLM: visual reasoning with a general conditioning layer, in \textit{Proceedings of the AAAI Conference on Artificial Intelligence}, Vol.~32 (2018).

\bibitem{Dan2018}
D. Mendels, G. M. Piccini, and M. Parrinello, Collective variables from local fluctuations, \doilink{10.1021/acs.jpclett.8b00733}{J. Phys. Chem. Lett. \textbf{9}, 2776 (2018)}.

\bibitem{Piccini2018MetadynamicsDiscriminants}
G. M. Piccini, D. Mendels, and M. Parrinello, Metadynamics with discriminants: a tool for understanding chemistry, \doilink{10.1021/acs.jctc.8b00634}{J. Chem. Theory Comput. \textbf{14}, 5040 (2018)}.

\bibitem{Mendels2018FoldingHLDA}
D. Mendels, G. M. Piccini, Z. F. Brotzakis, Y. I. Yang, and M. Parrinello, Folding a small protein using harmonic linear discriminant analysis, \doilink{10.1063/1.5053566}{J. Chem. Phys. \textbf{149}, 194113 (2018)}.

\bibitem{MendelsParrinello2019}
V. Rizzi, D. Mendels, E. Sicilia, and M. Parrinello, Blind search for complex chemical pathways using harmonic linear discriminant analysis, \doilink{10.1021/acs.jctc.9b00358}{J. Chem. Theory Comput. \textbf{15}, 4507 (2019)}.

\bibitem{Krueger2026IDPDesign}
R. K. Krueger, M. P. Brenner, and K. Shrinivas, Generalized design of sequence--ensemble--function relationships for intrinsically disordered proteins, \doilink{10.1038/s43588-025-00881-y}{Nat. Comput. Sci. \textbf{6}, 512 (2026)}.

\bibitem{Wasserstein}
M. Cuturi, Sinkhorn distances: lightspeed computation of optimal transport, in \textit{Advances in Neural Information Processing Systems}, Vol.~26 (2013), \href{https://arxiv.org/abs/1306.0895}{arXiv:1306.0895}.

\bibitem{adam}
D. P. Kingma and J. Ba, Adam: a method for stochastic optimization, in \textit{International Conference on Learning Representations (ICLR)} (2015), \href{https://arxiv.org/abs/1412.6980}{arXiv:1412.6980}.

\bibitem{bergstra2011}
J. Bergstra, R. Bardenet, Y. Bengio, and B. K\'egl, Algorithms for hyper-parameter optimization, in \href{https://proceedings.nips.cc/paper_files/paper/2011/file/86e8f7ab32cfd12577bc2619bc635690-Paper.pdf}{\textit{Advances in Neural Information Processing Systems}}, Vol.~24 (2011), pp.~2546--2554.

\bibitem{watanabe2023}
S. Watanabe, Tree-structured Parzen estimator: understanding its algorithm components and their roles for better empirical performance, \href{https://arxiv.org/abs/2304.11127}{arXiv:2304.11127}.

\bibitem{optuna}
T. Akiba, S. Sano, T. Yanase, T. Ohta, and M. Koyama, Optuna: a next-generation hyperparameter optimization framework, in \textit{Proceedings of the 25th ACM SIGKDD International Conference on Knowledge Discovery \& Data Mining} (2019), pp.~2623--2631, \doilink{10.1145/3292500.3330701}{doi:10.1145/3292500.3330701}.

\end{thebibliography}
\end{document}